\documentclass[onecolumn,showpacs,pre]{revtex4}
\usepackage{graphicx}
\usepackage{subfigure}

\begin{document}

\title{A model of a dual-core matter-wave soliton laser }
\author{Peter Y. P. Chen}
\affiliation{School of Mechanical and Manufacturing Engineering, University of New South
Wales, Sydney 2052, Australia}
\author{Boris A. Malomed}
\affiliation{Department of Interdisciplinary Studies, School of Electrical Engineering,
Faculty of Engineering, Tel Aviv University, Tel Aviv 69978, Israel}

\begin{abstract}
We propose a system which can generate a periodic array of
solitary-wave pulses from a finite reservoir of coherent
Bose-Einstein condensate (BEC). The system is built as a set of
two parallel quasi-one-dimensional traps (the reservoir proper and
a pulse-generating cavity), which are linearly coupled by the
tunneling of atoms. The scattering length is tuned to be negative
and small in the absolute value in the cavity, and still smaller
but positive in the reservoir. Additionally, a parabolic potential
profile is created around the center of the cavity. Both edges of
the reservoir and one edge of the cavity are impenetrable.
Solitons are released through the other cavity's edge, which is
semi-transparent. Two different regimes of the intrinsic operation
of the laser are identified: circulations of a narrow
wave-function pulse in the cavity, and oscillations of a broad
standing pulse. The latter regime is stable, readily providing for
the generation of an array containing up to $10,000$
permanent-shape pulses. The circulation regime provides for no
more than $40$ cycles, and then it transforms into the oscillation
mode. The dependence of the dynamical regime on parameters of the
system is investigated in detail.
\end{abstract}

\maketitle

\section{Introduction}

The concept of matter-wave lasers, which are intended to generate
strong coherent beams of atoms, and many applications that such
beams can find, are well known. It is commonly assumed that a
Bose-Einstein condensate (BEC) may serve as a source of the
matter-wave laser\ beam. Atomic lasing was considered in detail
theoretically (see Refs. \cite{laser} and references therein) and
demonstrated in various experimental settings \cite{experiment};
the topic was recently reviewed in Ref. \cite{review}. Of special
interest is a possibility to construct a laser operating in a
pulsed (quasi-soliton) regime, i.e., periodically releasing narrow
localized pulses of coherent atom waves, as proposed in Refs.
\cite{soliton-laser} and \cite{Carr}. In that connection, it is
relevant to mention that (effectively) one-dimensional solitons in
BEC with very weak attraction between atoms (in $^{7}$Li) were
created in well-known experiments \cite{soliton}.

The first models of the pulsed matter-wave lasers assumed release
of quasi-solitons from an elongated (``cigar-shaped") trap filled
by a self-attractive BEC. This scheme, while probably realizable
in the experiment, does not offer sufficiently effective means to
control the operation regime, and does not secure generation of a
very large number of pulses in a nearly periodic fashion. A more
sophisticated scheme was very recently proposed in Ref.
\cite{Spain}, which assumes zero nonlinearity coefficient (i.e.,
zero scattering length of inter-atomic collisions) in a larger
part of the elongated trap, and attraction between atoms (negative
scattering length)\ in its smaller part, where the pulses are to
be formed and released. The sign of the scattering length may be
controlled and altered along the length of the trap by means of
the Feshbach resonance\ \cite{Feshbach} (this mechanism was used
for the experimental creation of the BEC solitons \cite{soliton}).

In this work, we aim to propose a different model of a pulsed matter-wave
laser, in which the BEC reservoir and the pulse-generating cavity are
separated. As shown in Fig. \ref{setup} below, they are elongated parallel
traps with linear coupling between them, due to tunneling of atoms. Using
the aforementioned possibility to control the scattering length by means of
the Feshbach resonance, we assume that the interaction between atoms is
(very weakly) repulsive in the reservoir, while in the cavity it is
attractive (and weak too, see below). Additionally, the negative scattering
length is assumed to be altered along the cavity's axis, which can be
achieved using an appropriate configuration of the magnetic field
responsible for the Feshbach resonance (we do not assume any time dependence
of the scattering length or other parameters).

Besides using the magnetic field,\ it was predicted theoretically
\cite{optFRtheory} and demonstrated experimentally
\cite{optFRexperiment} that\ the Feshbach resonance can be
provided too by an adequately tuned optical field. Accordingly, a
spatially modulated stationary distribution of the light intensity
may also be employed to provide for necessary spatially
inhomogeneous Feshbach-resonance configurations in the reservoir
and cavity.

It is relevant to mention that various soliton solutions in dual-core traps,
with the linear coupling between the cores and \emph{opposite} signs of the
scattering length in them, were recently studied in detail \cite{Valery},
following an earlier work which was dealing with a model of dual-core
nonlinear optical fibers with opposite signs of the group-velocity
dispersion (rather than nonlinearity) in the two cores \cite{Dave}.

In the model introduced below, both edges of the reservoir, and
the left edge of the cavity are impenetrable to atoms, while the
right edge of the cavity (a ``valve")\ allows the release of
pulses into an outcoupling atomic waveguide, see Fig. \ref{setup}.
The model was originally designed with an intention to provide for
formation of a narrow quasi-soliton pulse in the cavity, that
would periodically circulate in it, bouncing from the edges. Each
time that the soliton hits the right edge (valve), a pulse is
released into the outcoupling waveguide. After being slashed this
way, the intrinsic solitary pulse is supposed to replenish itself
in the course of the subsequent cycle of the circulation by
collecting atoms tunneling from the reservoir. Accordingly, the
initial condition is taken with a large number of atoms in the
reservoir, and a small number in the cavity.

As described in detail below, numerical simulations demonstrate
that the circulation regime outlined above may be observed, but it
is not a really stable one: the intrinsic pulse would perform no
more than $40$ circulations, quickly coming to a halt and getting
broad. Nevertheless, a very robust regime of the periodic release
of pulses can be found in a large parameter region. In this
regime, a broad ``lump" stays immobile in the cavity, performing
periodic cycles of stretching and compression. When its right wing
reaches the valve, in the course of each cycle of the vibrations,
a new outcoupling pulse is released. The respective loss in the
number of atoms in the cavity is compensated by the influx of
atoms tunneling from the reservoir. The pulse-release cycles
repeat very many times, till the reservoir gets essentially
depleted. We stress that the model is a \emph{passive }one, i.e.,
it supports the stable pulse generation regime without any
externally applied active control, such as periodically opening
and shutting the valve by some clock signal.
\begin{figure}[tbp]
\includegraphics[width=4.00in]{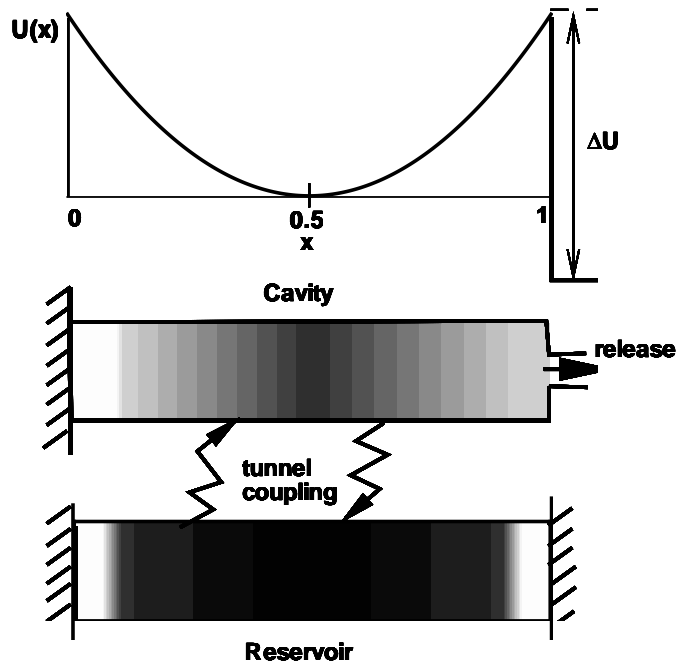}
\caption{A scheme of the proposed matter-wave laser composed of two nearly
one-dimensional traps coupled by tunneling of atoms, with negative and
positive scattering lengths in the cavity and reservoir, respectively. The
opening at the right edge of the cavity ($x=1$) depicts the valve releasing
pulses into an outcoupling waveguide. Shading symbolically shows an
instantaneous density distribution, $\left\vert \protect\psi \right\vert
^{2} $ and $\left\vert \protect\phi \right\vert ^{2}$, in both traps (larger
density corresponds to more intense shading). The upper part shows the
potential profile in the cavity, with the drop $\Delta U$ at $x=1$
corresponding to the valve.}
\label{setup}
\end{figure}

The paper is organized as follows. In Section 2, the model is formulated.
Basic results, which demonstrate both unstable circulation-based and stable
vibration-based regimes of the periodic generation of solitary pulses, are
presented in Section 3. In Section 4, we analyze the dependence of the
operation of the matter-wave laser on its parameters, and give an estimate
for the actual number of atoms in outcoupled pulses generated by the laser.
The paper is concluded by Section 4.

\section{The model}

According to what was said above, the model of the matter-wave laser is
based on a system of linearly coupled Gross-Pitaevskii equations (GPEs) for
the wave functions $\psi (x,t)$ and $\phi (x,t)$ of atoms in the
(effectively) one-dimensional cavity and reservoir, with opposite signs in
front of the nonlinear terms. We scale $\hbar $, the atomic mass $m$, and
the nonlinearity coefficient at $x=0$ in the cavity (in physical units, the
latter is $4\pi \hbar ^{2}a/m$, where $a$ is the $s$-wave scattering length
of atomic collisions around $x=0$) to be $1$. Then, the coupled equations
are cast in a normalized form,
\begin{eqnarray}
i\psi _{t} &=&-\frac{1}{2}\psi _{xx}-\left( 1+gx\right) |\psi |^{2}\psi
\nonumber \\
&&+f(x-L/2)^{2}\psi -\kappa \phi ,  \label{psi} \\
i\phi _{t} &=&-\frac{1}{2}\phi _{xx}+\epsilon |\phi |^{2}\phi -\kappa \psi ,
\label{phi}
\end{eqnarray}where the coordinate $x$ takes values $0<x<L$ (i.e., $L$ is the length of
both the cavity and reservoir). In accordance with what was said above, we
assume a small positive scattering length (weak repulsion between atoms) in
the reservoir, $\epsilon >0$, while in the cavity the interaction is
attractive, with the strength linearly increasing with $x$ ($g>0$ is
assumed), following a corresponding profile of the static magnetic or
optical field that controls the scattering length in the cavity via the
Feshbach resonance.

The coefficient $g$ will be small (see below), but it is needed to
create an effective potential slope in the cavity: as the
intra-cavity pulse gets ``heavier" due to the agglomeration of
atoms tunneling from the reservoir, the term proportional to $g$
creates a gradually increasing effective potential which pushes
atoms to the right (since $g$ is positive). In the aforementioned
circulation regime, this effective (nonlinear) potential pushes
the intra-cavity localized pulse to the right. After hitting the
right edge (at $x=L$), where the ``valve" is placed (see details
below), the pulse loses a part of its mass, generating the
outcoupled soliton. Therefore, the force induced by the nonlinear
potential drops, allowing the intra-cavity pulse to move in the
opposite direction after the bounce from the right edge, until it
will bounce (without loss) from the left edge, $x=0$. This way,
the nonlinear term proportional to $g$ may support periodic
circulations of the internal pulse (although, as said above, in
the domain of the model's parameter space explored in this work
the circulations are not really stable, eventually switching into
the regime of vibrations; the latter regime is possible with $g=0$
as well).

Further, we assume that an external parabolic potential acts in the cavity,
\begin{equation}
U_{\mathrm{pot}}(x)=f(x-L/2)^{2},  \label{Upot}
\end{equation}with the center set at the middle of the cavity. It is well known that such
a potential corresponds to a magnetic or optical trap applied to the
condensate \cite{optical}. It will be shown below that the results are not
sensitive to the exact shape of the intra-cavity potential; in fact,
essentially the same stable vibration regime can be obtained in the model
with no potential. \ Note that Eq. (\ref{phi}) for the reservoir does not
include any potential, as it is not necessary to generate the dynamical
regime sought for. We also explored a model including a potential in the
reservoir, but it did not affect the operation regime in any remarkable
manner.

The linear coupling between the cavity and reservoir is accounted
for by the positive coefficient $\kappa $, which determines the
tunneling time $1/\kappa $ for atoms. Keeping $\epsilon ,g,f$ and
$\kappa $ as free parameters of the model, the remaining scaling
invariance of Eqs. (\ref{psi}) and (\ref{phi}) makes it possible
to set $L\equiv 1$, which is fixed below.

As said above, the model assumes that the reservoir at its both edges, and
the cavity at its left edge are closed by impenetrable lids, which gives
rise to the respective boundary conditions (b.c.),
\begin{equation}
\phi (x=0)=\phi (x=1)=\psi (x=0)=0.  \label{0}
\end{equation}The pulses (solitons) are to be released into an external waveguide from the
cavity at its right edge, $x=1$ (recall we have set $L\equiv 1$).
For this purpose, a valve is set at $x=1$, assuming that the
corresponding b.c. for the $\psi $ field is linear. Actually, the
single possible form of such a linear b.c. is\begin{equation} \psi
_{x}(x=1)=iq\psi (x=1),  \label{tau}
\end{equation}with a positive constant $q$ (its physical meaning is discussed below).
Indeed, using the continuity equation for the total density, $\rho
\equiv |\psi |^{2}+|\phi |^{2}$, and current, $j=(i/2)\left( \psi
_{x}^{\ast }\psi -\psi _{x}\psi ^{\ast }\right) +(i/2)\left( \phi
_{x}^{\ast }\phi -\phi _{x}\phi ^{\ast }\right) \equiv
j_{1}+j_{2}$, in the coupled GPEs (\ref{psi}) and (\ref{phi}),
\begin{equation}
\frac{\partial \rho }{\partial t}=\frac{i}{2}\left( \psi _{xx}\psi ^{\ast
}-\psi _{xx}^{\ast }\psi \right) \equiv -\frac{\partial j}{\partial x},
\label{rho}
\end{equation}integrating Eq. (\ref{rho}) over the length of the system, $0<x<1$, and
taking into regard that $j_{1}(x=0)=j_{2}(x=0)=j_{2}(x=1)=0$
pursuant to b.c. (\ref{0}), one can derive a \textit{depletion
equation} for the solution's norm $N$, which is proportional to
the total number of atoms in the system:\begin{eqnarray} N
&=&\int_{0}^{1}\left( |\psi (x)|^{2}+|\phi (x)|^{2}\right)
dx\equiv
N_{1}+N_{2},  \label{N} \\
-\frac{dN}{dt} &=&j_{1}(x=1,t)=q\left\vert \psi (x=1,t)\right\vert
^{2}\equiv E(t),  \label{E}
\end{eqnarray}where b.c. (\ref{tau}) was substituted for $\psi _{x}$ in the expression for
$j_{1}(x=1)$. The latter result is what might be expected: the
instantaneous rate at which the density is released through the
valve is proportional to the local density $\left\vert \psi
(x=1)\right\vert ^{2}$, which justifies the adoption of b.c.
(\ref{tau}). Initial values of the norms $N_{1}^{(0)}$ and
$N_{2}^{(0)}$ of the $\psi $ and $\phi $ fields, and the b.c.
constant $q $ are supplementary control parameters of the model,
in addition to the above-mentioned set of $\left( \epsilon
,g,f,\kappa \right) $. Below, it will be demonstrated that really
important parameters are $\epsilon $, $\kappa $, and $q$.

The physical meaning of b.c. (\ref{tau}) can be readily
understood. Indeed, it implies that the local gradient of the
phase of the wave function $\psi $ at the point $x=1$, which is
proportional to the atom's momentum, is fixed to be $q$. This can
be realized by assuming that, to the right of the point $x=0$ (in
the outcoupling waveguide), the cavity potential drops by $\Delta
U=q^{2}/2$ (recall we have set $m=\hbar =1$). The drop $\Delta U$
must be essentially larger than the potential and kinetic energy
of atoms to the left of the point $x=1$ (in the cavity). The
latter condition, which amounts to\begin{equation} U(x=1)=f/4\ll
q^{2}/2  \label{condition}
\end{equation}[see Eq. (\ref{Upot})], provides for the possibility to neglect small
deviations in the kinetic energy of the released atoms from $q^{2}/2$. We
notice that a soliton-releasing valve in the matter-wave laser model
recently proposed in Ref. \cite{Spain} includes a similar element (an
effective potential step).

The form of the function $E(t)$, as defined in Eq. (\ref{E}), determines the
shape of the density pattern (actually, an array of pulses) released into
the outcoupling waveguide, the same way as the temporal shape of the input
signal at the point $z=0$ determines the shape of a temporal soliton or
solitonic array propagating along the coordinate $z$ in the optical fiber
\cite{Agrawal}. Note that, as follows from Eq. (\ref{tau}), the limit cases
of $q=0$ and $q=\infty $ correspond, respectively, to a reflecting mirror
and impenetrable lid, i.e., $\psi _{x}(x=1)=0$ and $\psi (x=1)=0$. In either
case, the system does not release anything. In addition to yielding the
final result in the form of $E(t)$, Eq. (\ref{E}), with $dN/dt$ computed
directly from numerical data, provides for a means to monitor the accuracy
of simulations.

Finally, we note that the model does not include an explicit form of the
outcoupling waveguide (the extension of the cavity beyond $x=1$), which is
responsible for the eventual shaping the released pulses into solitons.
Actually, the soliton-formation problem in a uniform waveguide has been
studied in detail before (see, e.g., Ref. \cite{Carr} and references
therein).

\section{Operation regimes of the pulsed matter-wave laser}

In the numerical analysis of the model, Eqs. (\ref{psi}) and
(\ref{phi}) were solved by means of a finite-element
pseudospectral method, which implies division of the integration
domain into several finite elements. Solution in each element is
approximated by a polynomial, and a pseudo-spectral method, based
on Chebyshev collocation points \cite{Peter}, is used in each
element. We have checked that a numerical algorithm based on
domain stretching, rather than domain division into finite
elements \cite{Chebyshev}, produces the same results.

In cases when the stationary version of the GPEs, i.e., a system of two
linearly coupled ordinary differential equations, had to be solved (see
below), the finite-element pseudospectral method reduces them to a set of
algebraic equations, which were treated by means of the Newton's method. In
the general case, when Eqs. (\ref{psi}) and (\ref{phi}) are partial
differential equations, the time derivatives in the equations were
approximated by the finite difference as per the Crank-Nicholson scheme.
This, together with the finite-element pseudospectral method, yields a set
of nonlinear algebraic equations, that were also solved by the Newton's
algorithm.

In all the cases considered, the integration domain $0\leq x\leq
1$ was divided into four equal elements, with a fifth-order
polynomial used in each one. The procedure leads to a set of $48$
algebraic equations, and it was checked that increase of the
polynomial's order did not change the results in any tangible
aspect. We used the time step of $\Delta t=0.001$, checking that
smaller time steps gave virtually the same results (while larger
$\Delta t$ would generally be insufficient to produce accurate
results).

\begin{figure}[tbp]
$\begin{array}{cc}
\includegraphics[width=3.7in]{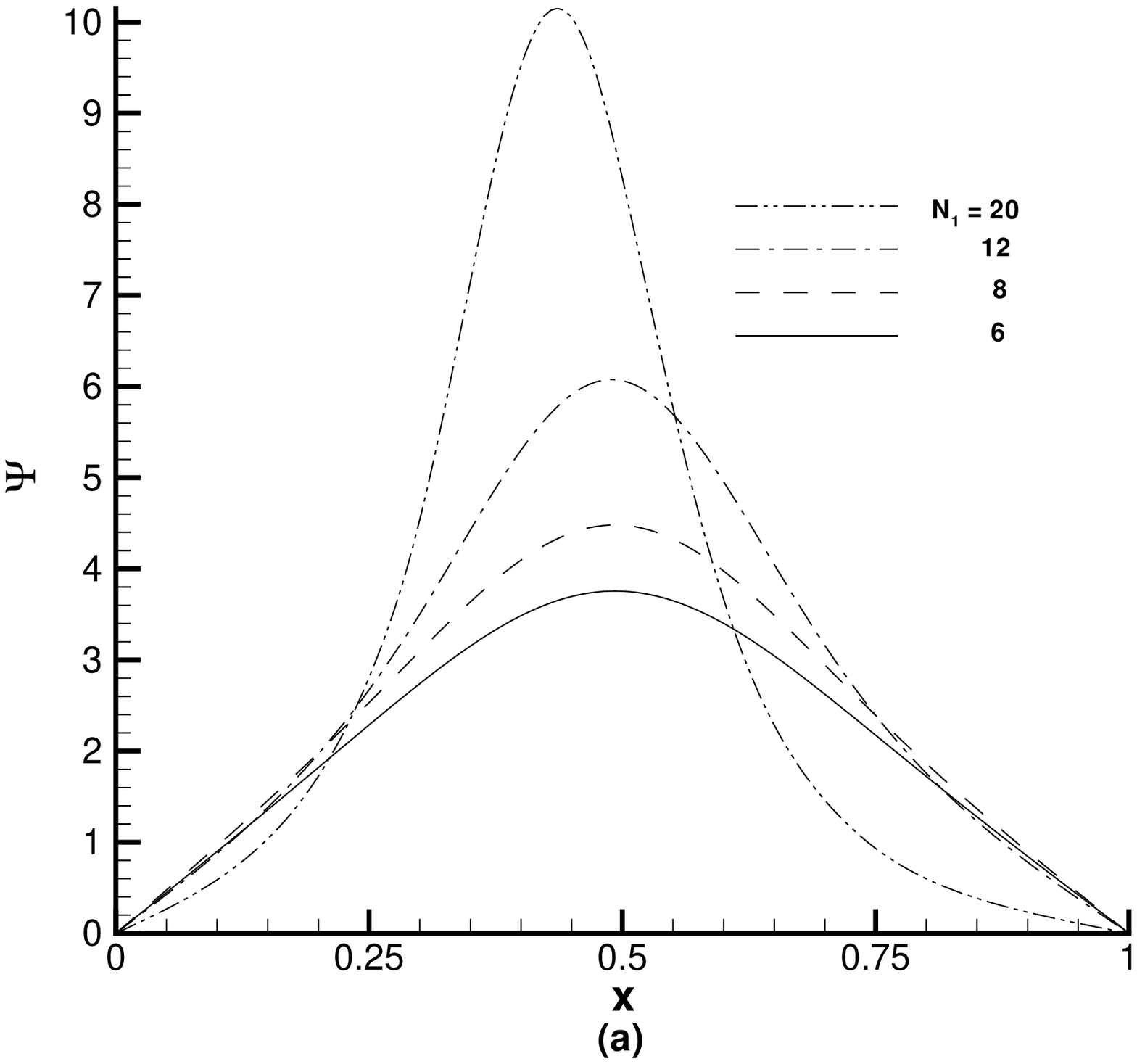} &
\includegraphics[width=3.7in]{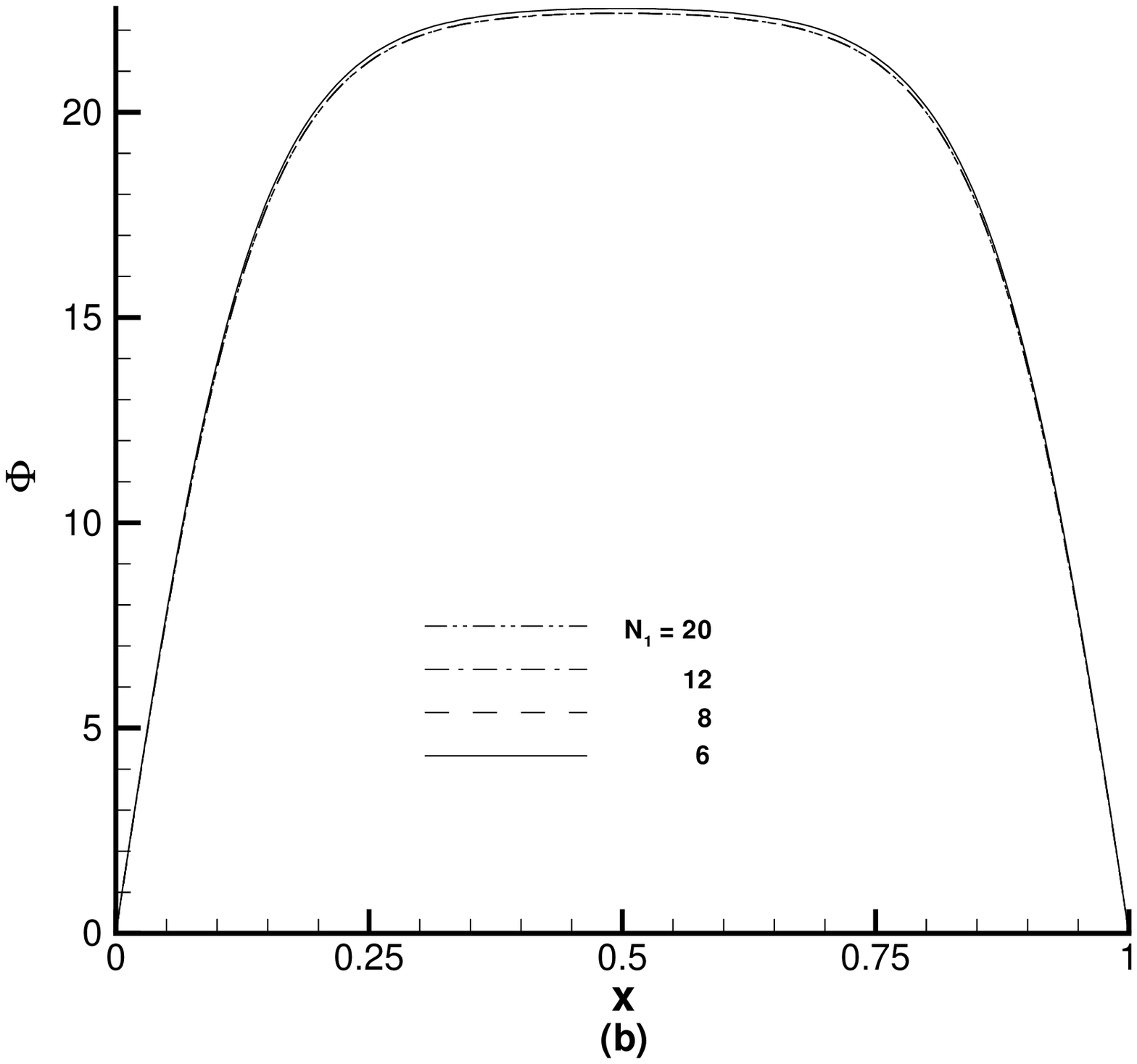}\end{array}$
\caption{Typical examples of the initial real profiles of $\Psi (x)$ (a) and
$\Phi (x)$ (b) for $g_{1}=0.02$, $\protect\epsilon =0.1$, and
$\protect\kappa =0.5$, used in simulations of the coupled
equations (\protect\ref{psi}) and (\protect\ref{phi}). In all the
cases, the norm of the $\protect\phi $-component is fixed (in the
normalized units) to be $N_{2}=363.4$, while the norm $N_{1}$ of
the $\protect\psi $-component varies as shown in the figure.}
\label{fig1}
\end{figure}

Simulations started with initial conditions which, by themselves,
were stationary solutions to Eqs. (\ref{psi}) and (\ref{phi}), in
the form of $\left\{ \psi (x,t),\phi (x,t)\right\} =\exp \left(
-i\mu t\right) \left\{ \Psi (x),\Phi (x)\right\} $ with real $\Psi
(x)$ and $\Phi (x)$. These stationary solutions were found, in a
numerical form, for given values of $g,\epsilon $ and $\kappa $.
While doing so, b.c. (\ref{tau}) at $x=1$ was temporarily replaced
by $\psi =0$ [as said above, this formally corresponds to
$q=\infty $ in Eq. (\ref{tau})], since a stationary solution is
not possible otherwise, due to the nonzero current $j_{1}$ across
the point $x=1$, see Eq. (\ref{rho}). Further, both equations
(\ref{psi}) and (\ref{phi}) used to generate the stationary
solutions were modified by including \textit{ad hoc} potentials
linear in $x$, as the so generated initial configurations were
found to quickly produce stable operation regimes, even though the
actual potentials in Eqs. (\ref{psi}) and (\ref{phi}) are
different [recall that Eq. (\ref{phi}) has no potential at all]. A
set of typical initial configurations obtained this way is
displayed in Fig. \ref{fig1}.

Starting from such initial configurations, simulations produced two types of
dynamical regimes. The first one features a relatively narrow pulse which
performs shuttle motion, circulating in the cavity. Each collision of the
pulse with the valve at the right edge of the cavity gives rise to a
quasi-soliton released by the matter-wave laser. However, as mentioned
above, the circulation regime was never found to be stable (although we
cannot claim that the investigation of the system's parameter space was
exhaustive, as it is difficult to perform complete exploration of the
seven-paramater space). The pulse would perform no more than $40$
circulations (typically, fewer), and this regime would transform itself into
a stable vibration regime.

\begin{figure}[tbp]
$\includegraphics[width=4.00in]{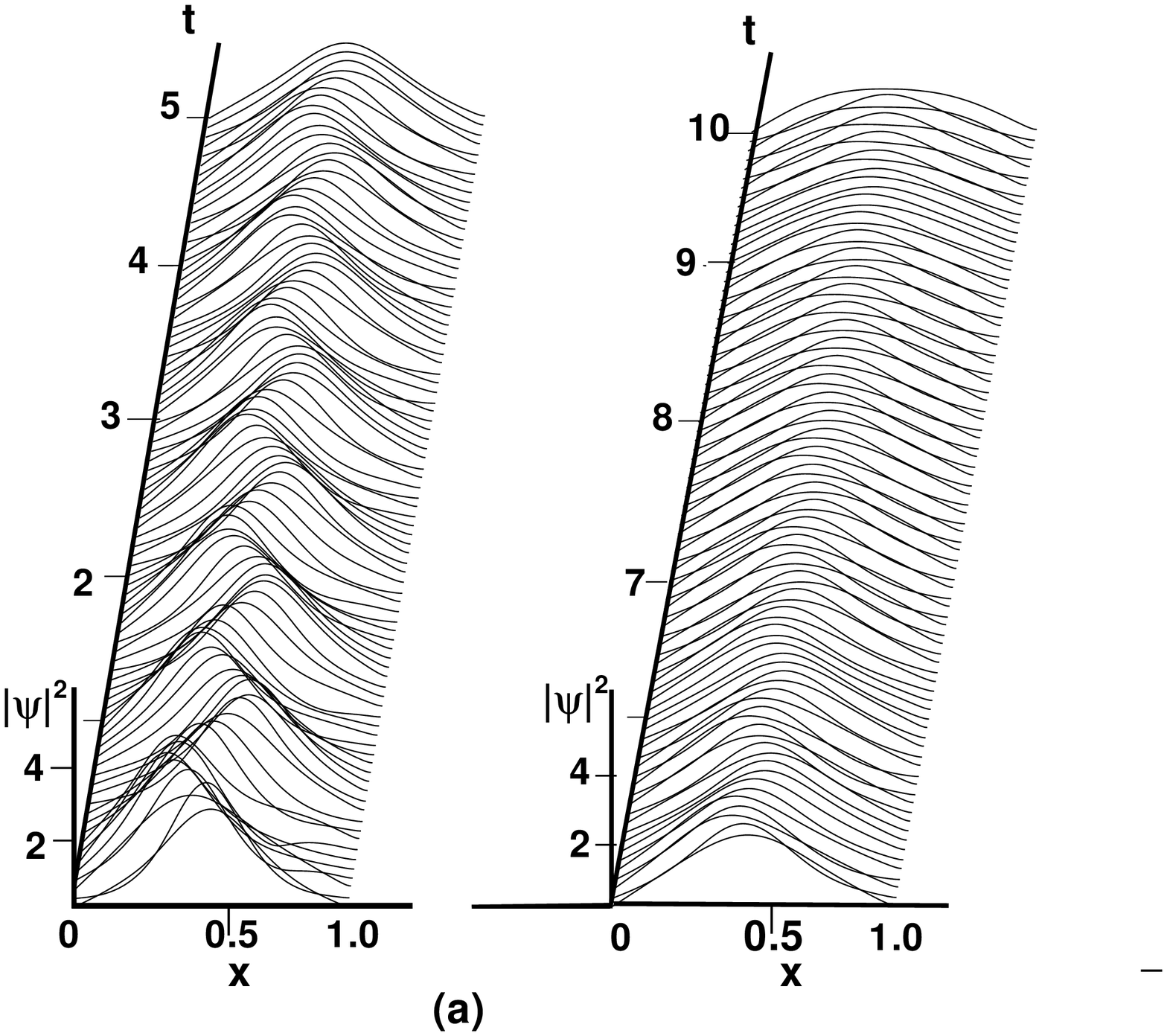}$ \newline
$\begin{array}{cc}
\includegraphics[width=3.7in]{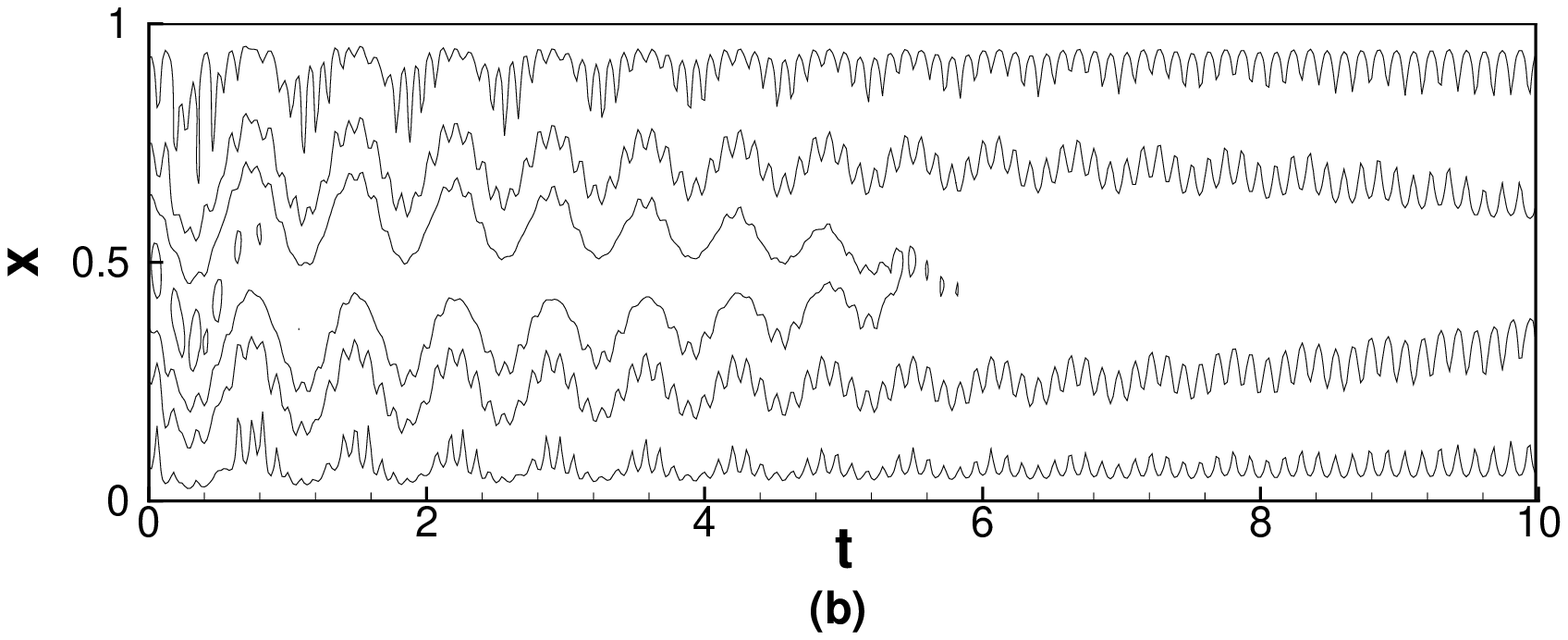} &
\includegraphics[width=3.7in]{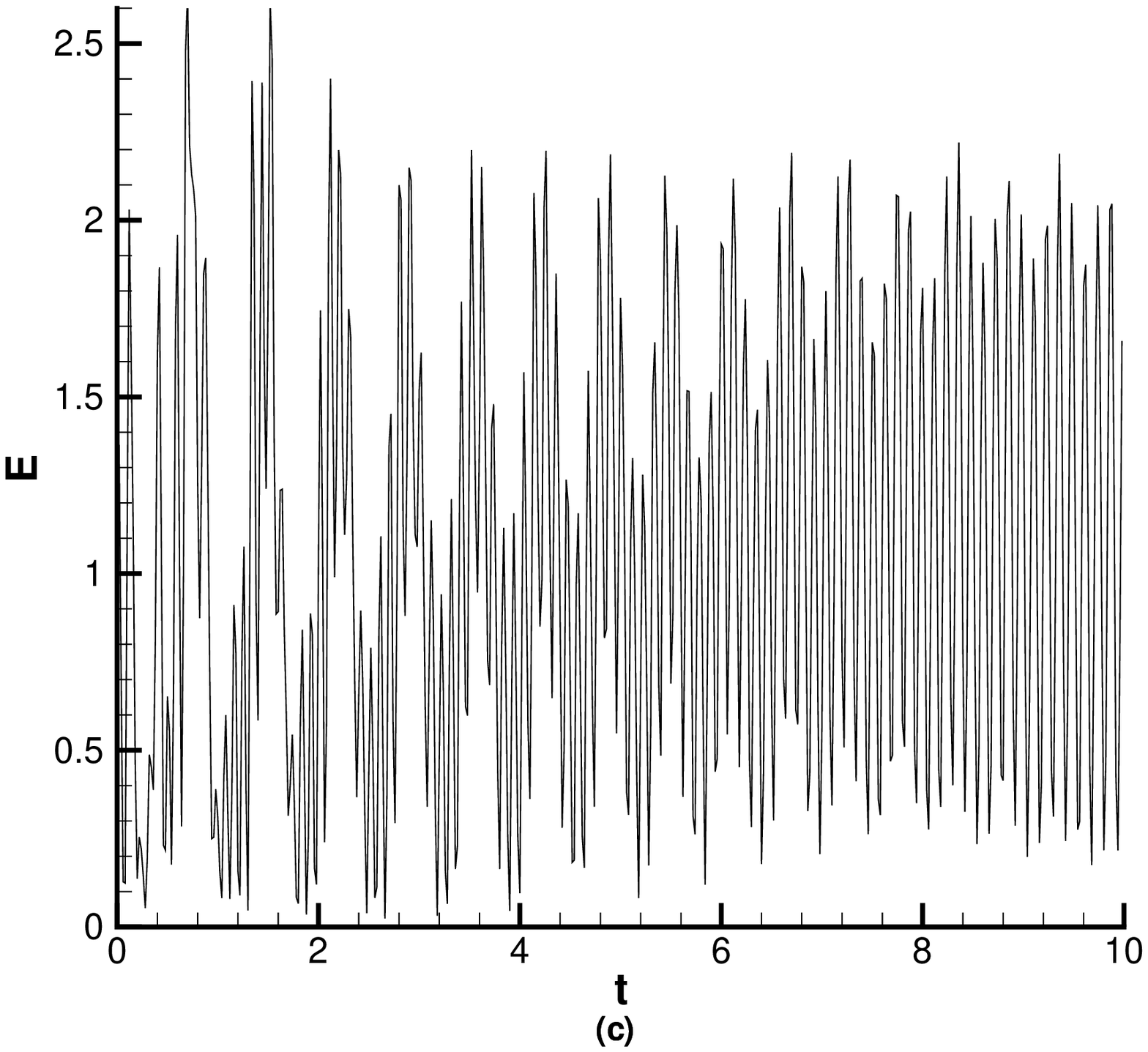}\end{array}$
\caption{Transition of the circulation regime into a vibration one, shown in
terms of a three-dimensional plot for the density $|\protect\psi
(x,t)|^{2}$ in the cavity (a), contour plots of the density (b),
and the outcoupling rate $E(t)$ (c). The parameters are $g=0.02$,
$\protect\epsilon =0.1$, $f=3$, $\protect\kappa =0.5$, and $q=50$.
The initial (normalized) numbers of atoms in the reservoir and
cavity are $N_{2}^{(0)}=360$ and $N_{1}^{(0)}=12$. } \label{fig2}
\end{figure}

An example of the transition from circulations to vibrations is displayed in
Fig. \ref{fig2}. Actually, the transition may be much shorter at other
values of the parameters. In most cases, as shown below, the vibration
regime sets in directly from the initial configuration, without going
through the transient stage of circulations.

The instability of the circulation regime can be understood. Indeed, one may
consider a small fluctuation which makes the norm of the pulse released into
the outcoupling waveguide slightly smaller than in the established regime,
hence the norm remaining in the cavity soliton becomes slightly larger.
Consequently, the strength of the effective nonlinear potential, that tries
to push the soliton to the right, drops (as a results of the bounce from the
valve) by an amount which is a bit smaller than in the unperturbed state,
and the soliton will therefore slide to the left slower than in the absence
of the perturbation. As a result, spending more time in the cavity, the
soliton will absorb more atoms in the course of the next cycle of the
circulations, and will return to the right edge with a still larger norm.
The mechanism by which a fluctuation slightly increasing the soliton's norm
leads to its still larger increase (or, conversely, a fluctuation decreasing
the norm causes its further decrease) implies the instability. It may happen
that the circulation regime can be stabilized in a model including an active
element (i.e., \textit{forced circulations} might be stable), but in this
work we focus on the more fundamental passive model.

\begin{figure}[tbp]
$\begin{array}{cc}
\includegraphics[width=3.7in]{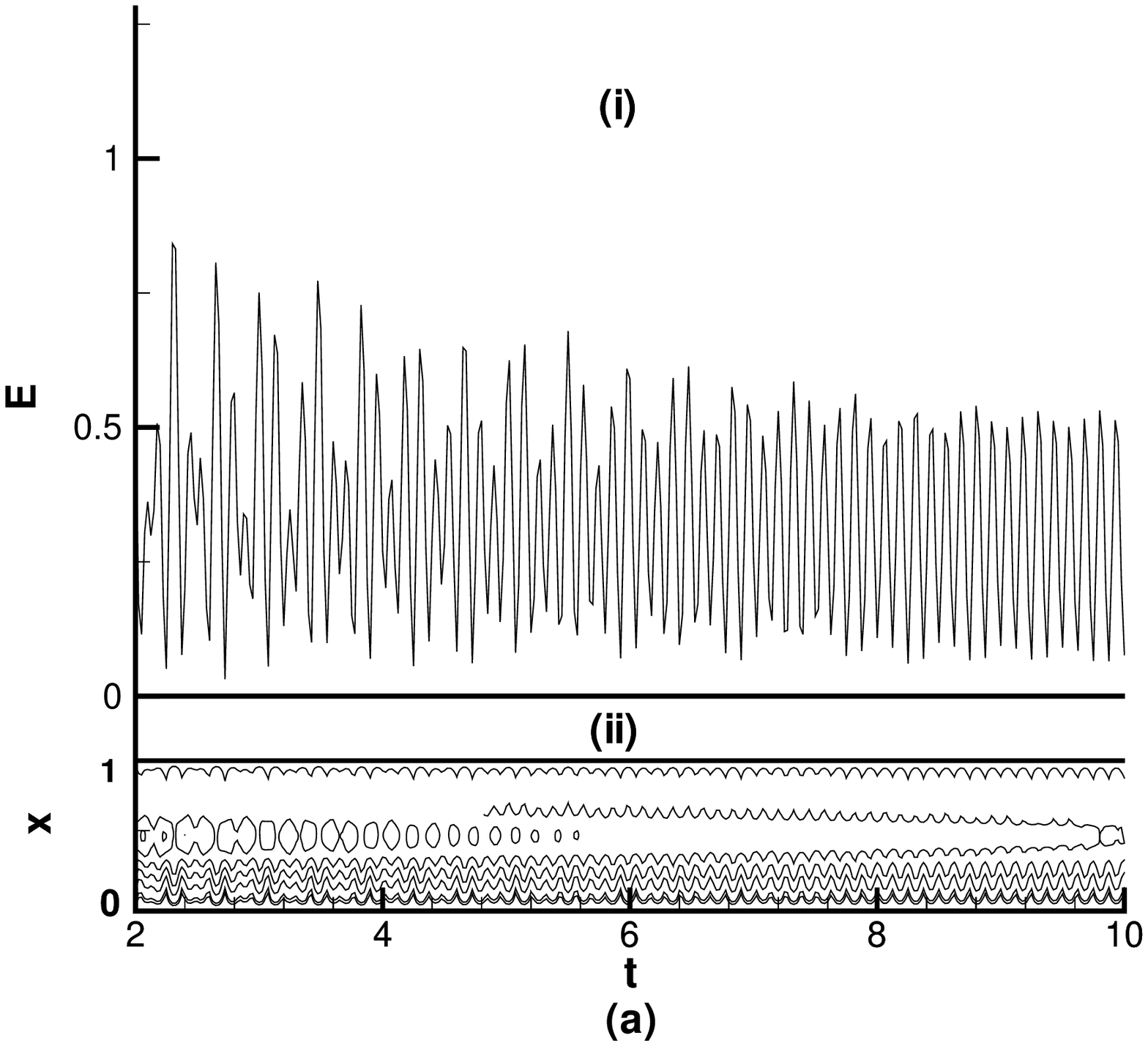} & \includegraphics[width=3.7in]{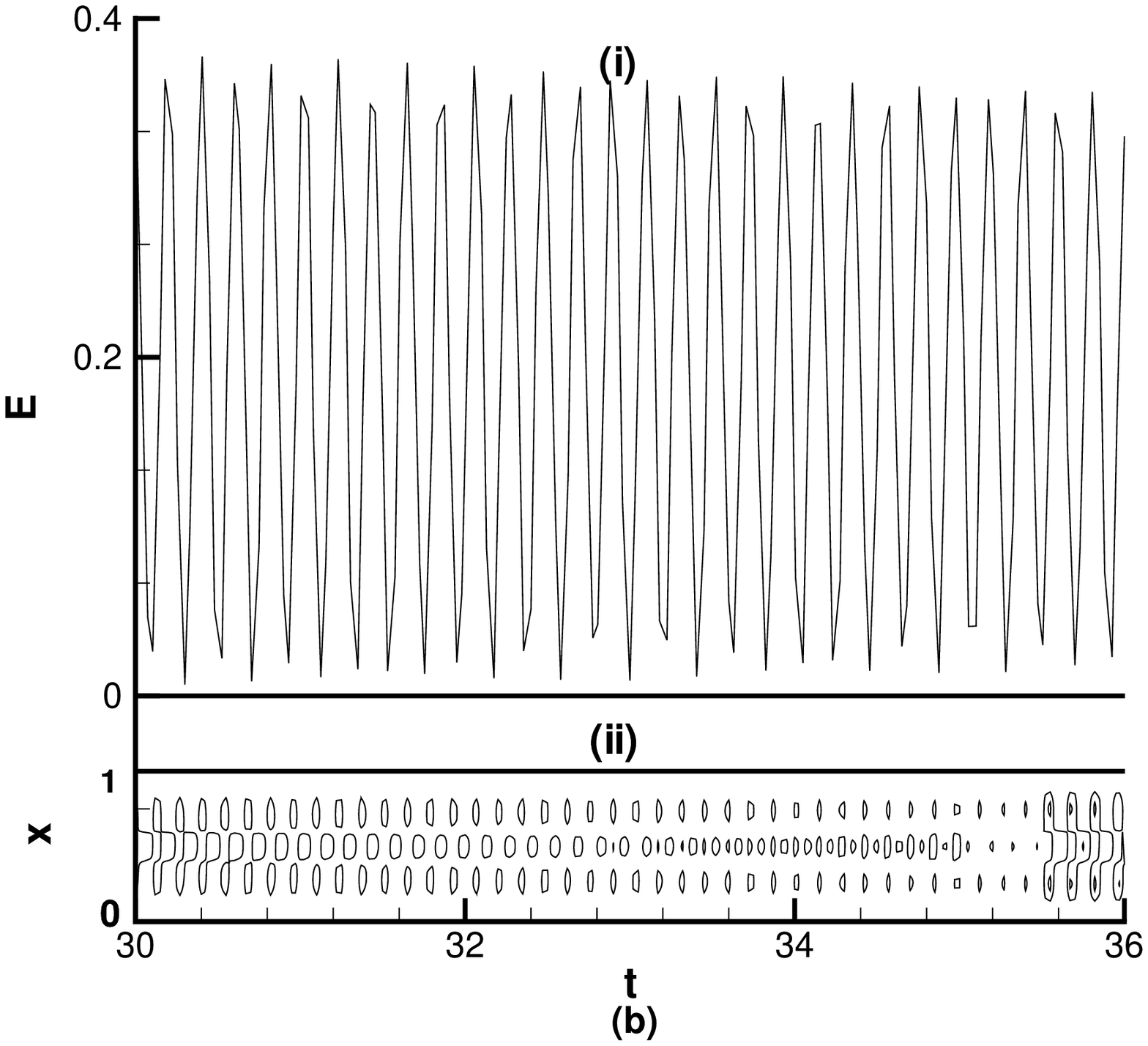} \\
\includegraphics[width=3.7in]{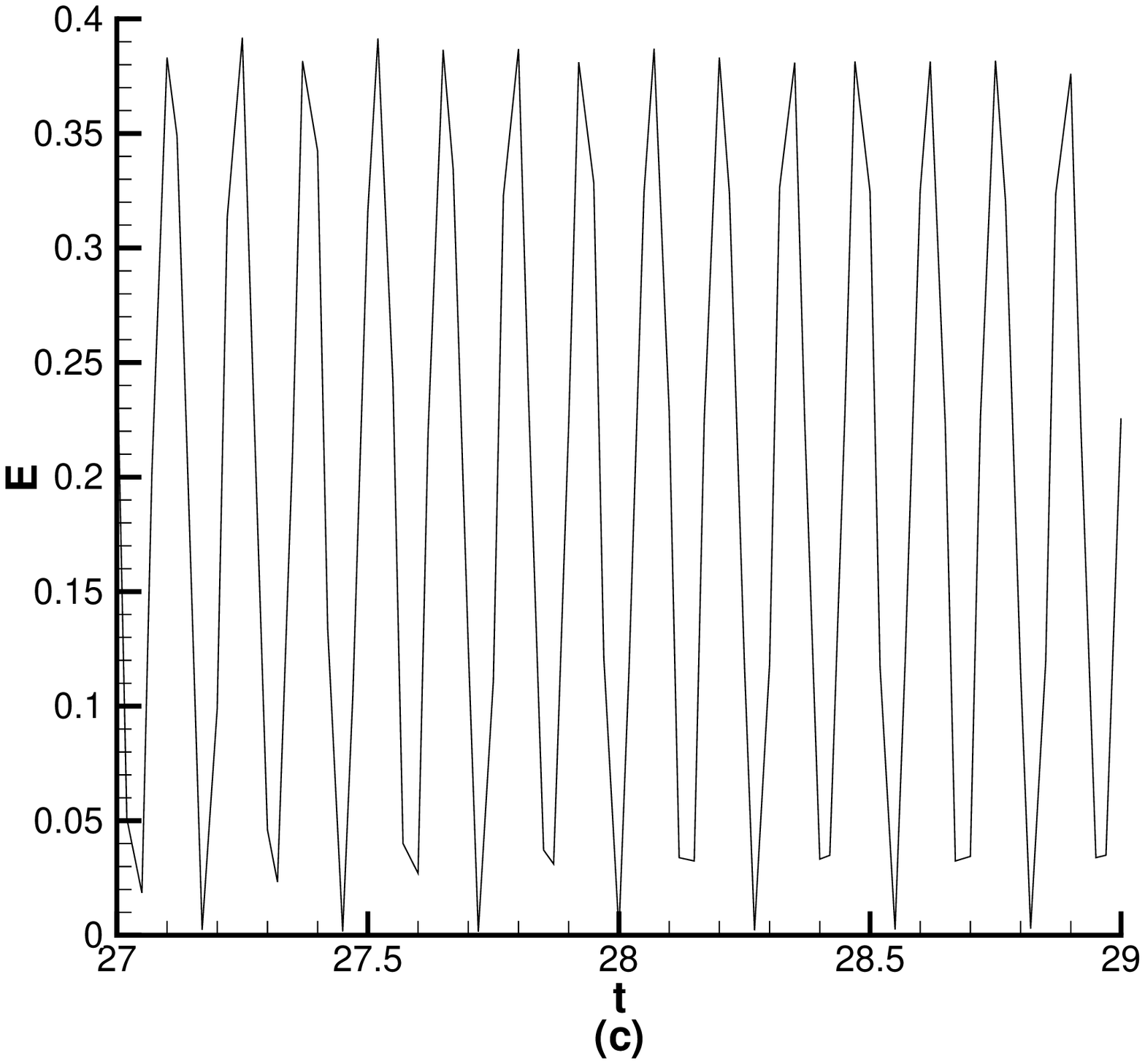} & \includegraphics[width=3.7in]{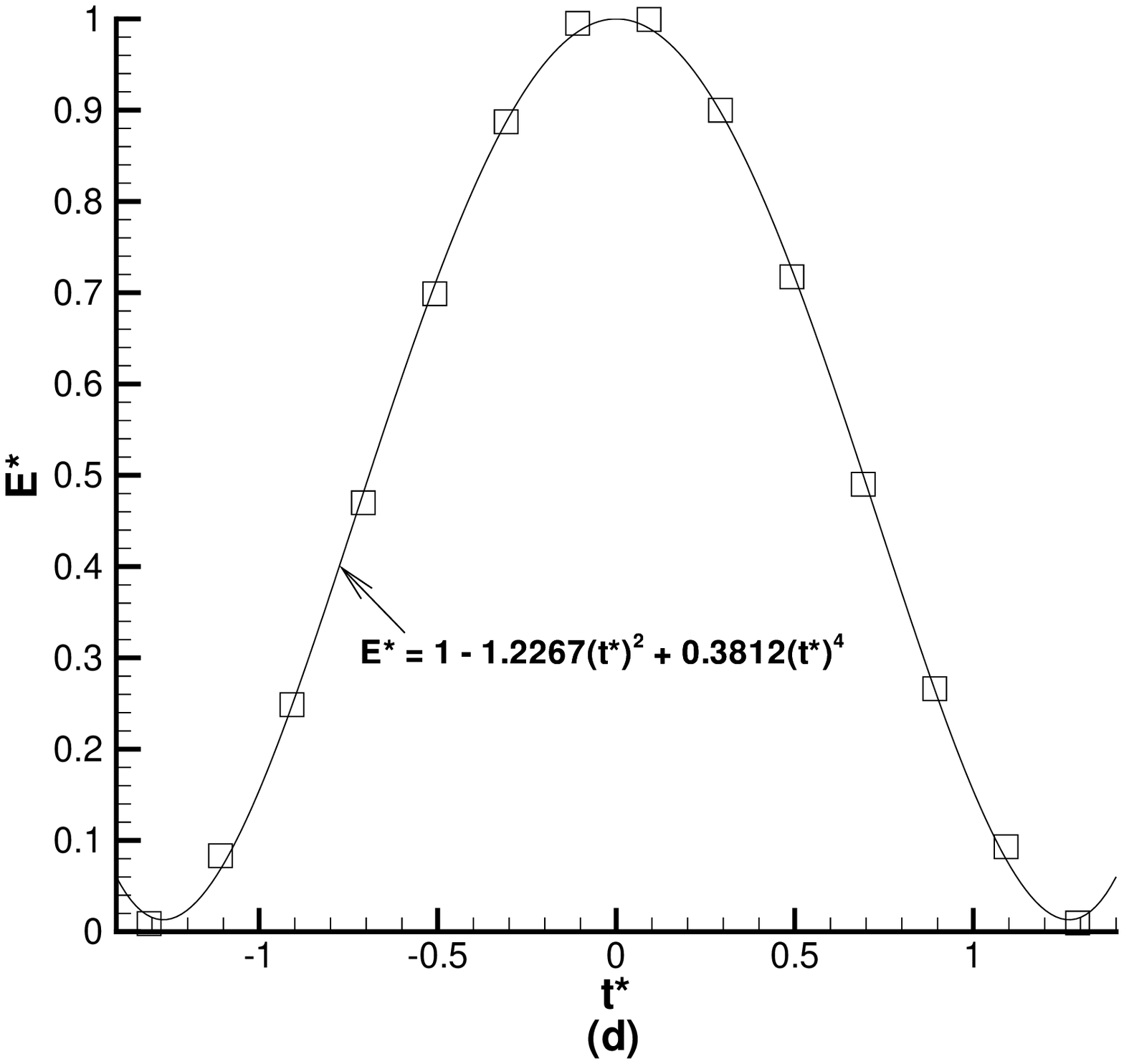}\end{array}$
\caption{A typical example of the establishment of a stable vibration regime
in the cavity, shown in terms of the outcoupling rate $E(t)$,
which determines the shape of the pulse array generated by the
matter-wave laser [$E$ is defined as per Eq. (\protect\ref{E})].
(a) The initial stage; (b) beginning of the established regime;
(c) a fragment from panel (b), showing 15 cycles in detail; (d) a
single typical cycle (shown in terms of obviously rescaled
variables $t^{\ast }$ and $E^{\ast }$), together with a simplest
analytical fit to it. In panels (a) and (b), contour plots in the
lower subpanels (ii) additionally illustrate the spatiotemporal
evolution of the atom density, $\left\vert \protect\psi
(x,t)\right\vert ^{2}$. Parameters are $g=0.02$, $\protect\epsilon
=0.10$, $f=3$, $\protect\kappa =0.5$, and $q=200$. The initial
normalized amounts of matter in the reservoir and cavity are
$N_{2}=363.4$ and $N_{1}=6$. In the established regime, the period
is $T=0.135$, with the duty cycle of $50\%$.} \label{fig3}
\end{figure}

\begin{figure}[tbp]
\includegraphics[width=4.00in]{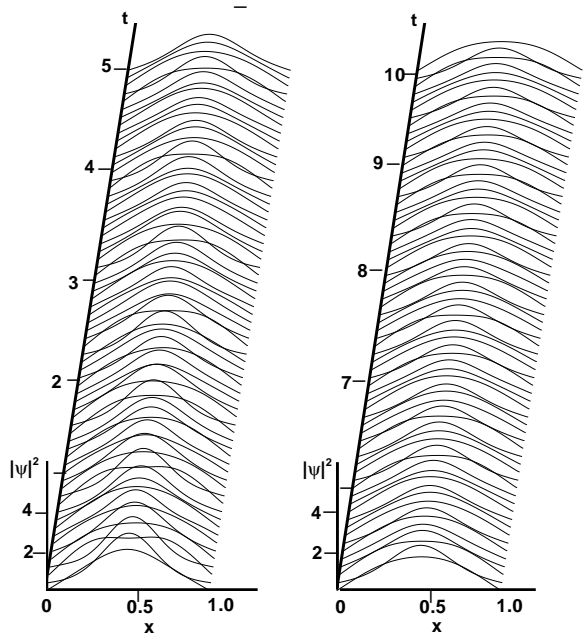}
\caption{The evolution of the density distribution in the cavity,
$|\protect\psi (x,t)|^{2}$, for the same case as shown in Fig.
\protect\ref{fig3}.} \label{fig4}
\end{figure}

In the regime of vibrations, a broad pulse stays at the center of
the cavity and performs a very large number of internal
oscillations. At a stage of the vibration cycle when the pulse
spreads out, its right wing hits the valve and generates an
outcoupling localized matter-wave packet. A typical example of
such a stable regime is displayed in Figs. \ref{fig3} and
\ref{fig4}. The second panel in Fig. \ref{fig3} shows (for a long
interval of time) the temporal dependence of the outcoupling rate
$E(t)$, as defined by Eq. (\ref{E}). As said above, this
dependence actually determines the shape of the pulse array
released by the matter-wave laser.

Figure \ref{fig4} depicts vibrations of the density distribution
in the cavity, for the same case as displayed in Fig. \ref{fig3}.
As seen from these figures, after a transient period, a very
robust regime (with some residual long-period modulations) sets
in. With the maximum value of the release rate $E$ in this regime
$E_{\max }\simeq 0.4$, the oscillation period $T=0.135$ and the
\textit{duty cycle} (the share of the period within which the
value of the density at the right edge of the cavity, $|\psi
(x=1,t)|^{2}$, exceeds half of its maximum value) being $0.5$
imply that the number of cycles which are expected before
depletion of the reservoir will become appreciable (say, $N$ will
drop to two thirds of $N_{2}^{(0)}=363.4$) can be easily
estimated:\begin{equation} \frac{(2/3)N_{2}^{(0)}}{E_{\max }\cdot
(T/2)}\simeq 10,000\text{ \textrm{cycles}.}  \label{cycles}
\end{equation}Direct verification of this prediction requires too long simulations
(up to $t\simeq 1500$). An estimate for the actual number of atoms in the generated
pulses (which may range between $10$ and $1000$) is given in the next
section.

It is relevant to stress that, in the cases presented in Figs.
\ref{fig2} and \ref{fig3}, \ref{fig4}, with $f=3$ and $q=200$, the
condition (\ref{condition}), which substantiates the use of b.c.
(\ref{tau}), is satisfied with a huge margin. In fact, this
condition holds equally well in all the cases where stable
operation of the matter-wave laser model was observed. It is also
relevant to mention that, in physical units, the value of $q\sim
100$ corresponds, for $^{7}$Li, to the velocity $\sim 1$ cm/s of
the released atoms, the respective kinetic energy being $\sim 0.1$
nK, on the temperature scale.

\begin{figure}[tbp]
\includegraphics[width=4.00in]{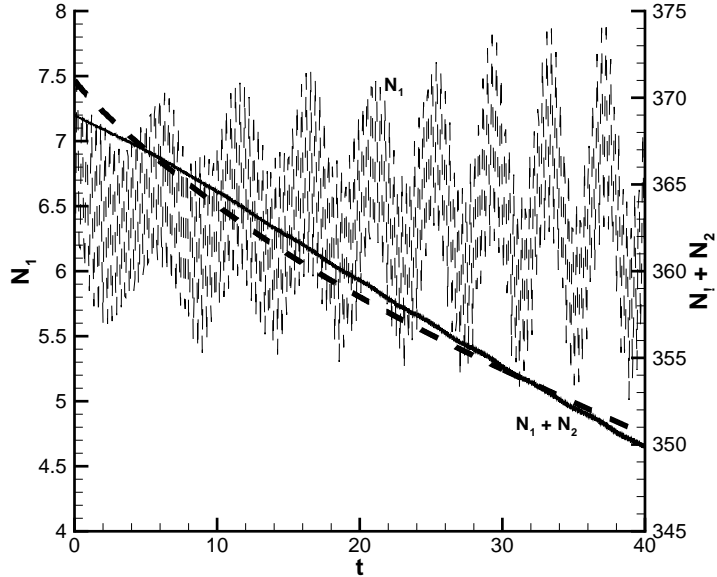}
\caption{The evolution of the norm (amount of matter) in the
reservoir, $N_{1}$ (the rapidly oscillating dashed curve), and the
total norm, $N_{1}+N_{2}$ (the continuous finite-thickness line),
in a stable operation regime. The bold dashed curve displays the
prediction following from the depletion equation
(\protect\ref{E}), averaged over rapid oscillations of $E(t)$.
Parameters are the same as in Fig. \protect\ref{fig3}.}
\label{fig5}
\end{figure}

It is relevant to compare the depletion rate in the stable operation mode
[as given by Eq. (\ref{E})] with the rate at which the cavity and reservoir
exchange the matter. A generic example of this comparison is displayed in
Fig. \ref{fig5}. As is seen, the exchange rate is much higher than the speed
of depletion. It is interesting to note that the long-period beatings in the
dependence of $N_{1}(t)$, which are observed in \ref{fig5}, practically do
not manifest themselves in the oscillations of the outcoupling rate $E(t)$
[nor in the evolution of $N_{1}+N_{2}$, as the beatings in $N_{1}(t)$ are
almost exactly compensated by anti-phase beatings in $N_{2}(t)$]. In other
words, the beatings do not affect the quality of the pulse array generated
by the matter-wave laser.

The line showing $N_{1}+N_{2}$ vs. $t$ in Fig. \ref{fig5} seems to
have finite thickness due to fast oscillations of $E(t)$, such as
those in Fig. \ref{fig3}. Its comparison with the respective
dependence predicted by averaging of the depletion equation
(\ref{E}) (a bold dashed line) is also shown in Fig. \ref{fig5}.
An error due to the averaging may account for a small discrepancy
with the actual evolution of $N_{1}+N_{2}$.

\section{Dependence of the operation regime on parameters}

The matter-wave laser model proposed above depends on several
parameters. It is necessary to identify those which are really
important to the stability of the operation mode, and which are
not. First of all, simulations demonstrate that there is virtually
no dependence on the potential's strength $f$ in Eq. (\ref{psi}):
at least within the interval of $1<f<3$, all dynamical
characteristics of the matter-wave laser\ model display very
little variation, if other parameters are fixed (taking typical
values of physical parameters given below, we conclude that, in
this parameter region, the range of $1<f<3$ actually corresponds
to the range of the longitudinal trapping frequencies $0.1-1$ Hz).
The dependence on the initial amount of matter in the cavity,
$N_{1}^{(0)}$, is very weak too, at least in the interval of
$5<N_{1}^{(0)}<20$. Also negligible is sensitivity of the
established regime to the parameter $g$ in Eq. (\ref{psi}). In
fact, additional simulations have demonstrated that virtually the
same stable oscillation regime sets in if we take Eqs. (\ref{psi})
and (\ref{phi}) with $f=g=0$.

\begin{figure}[tbp]
$\begin{array}{cc}
\includegraphics[width=3.7in]{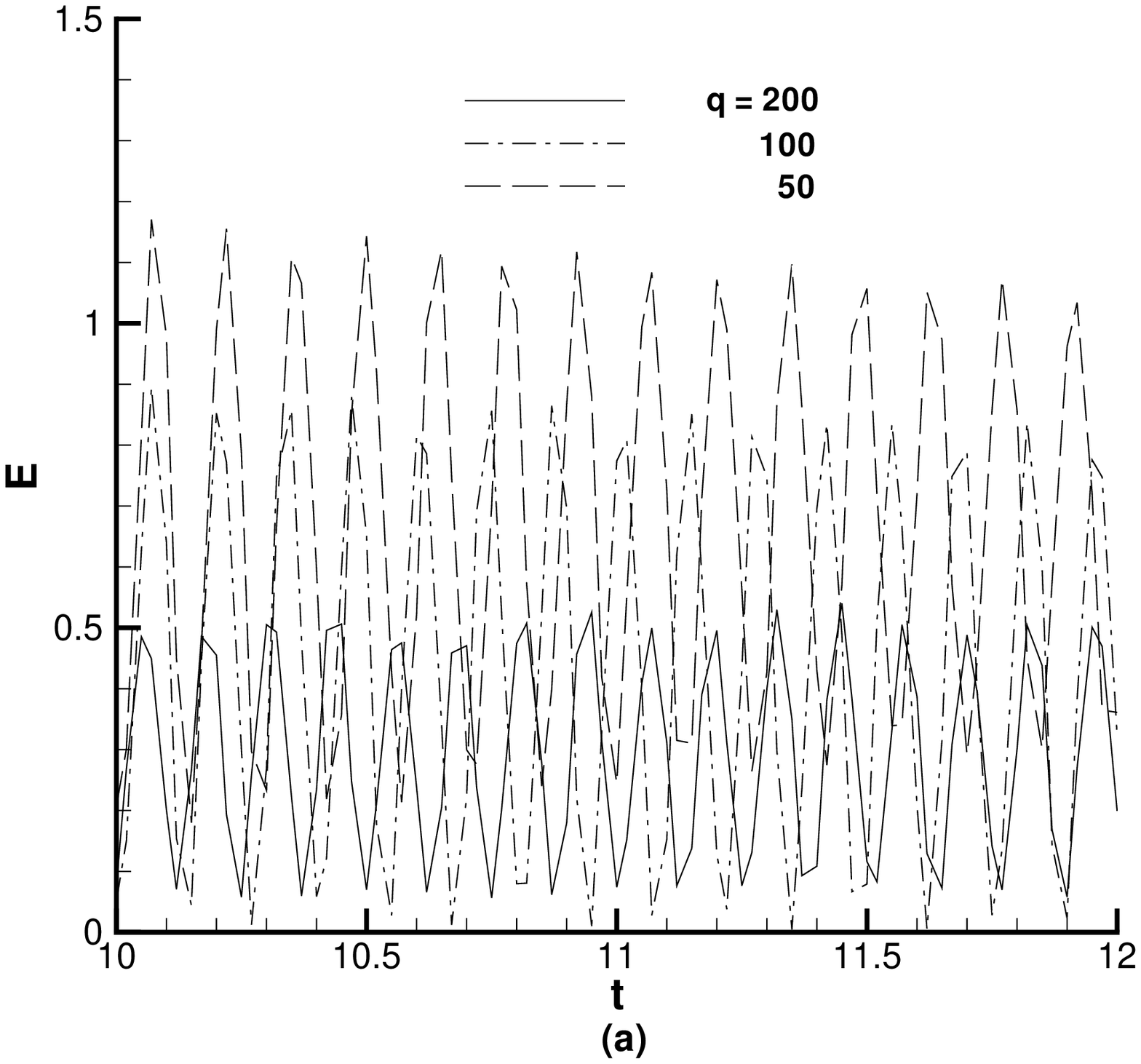} & \includegraphics[width=3.7in]{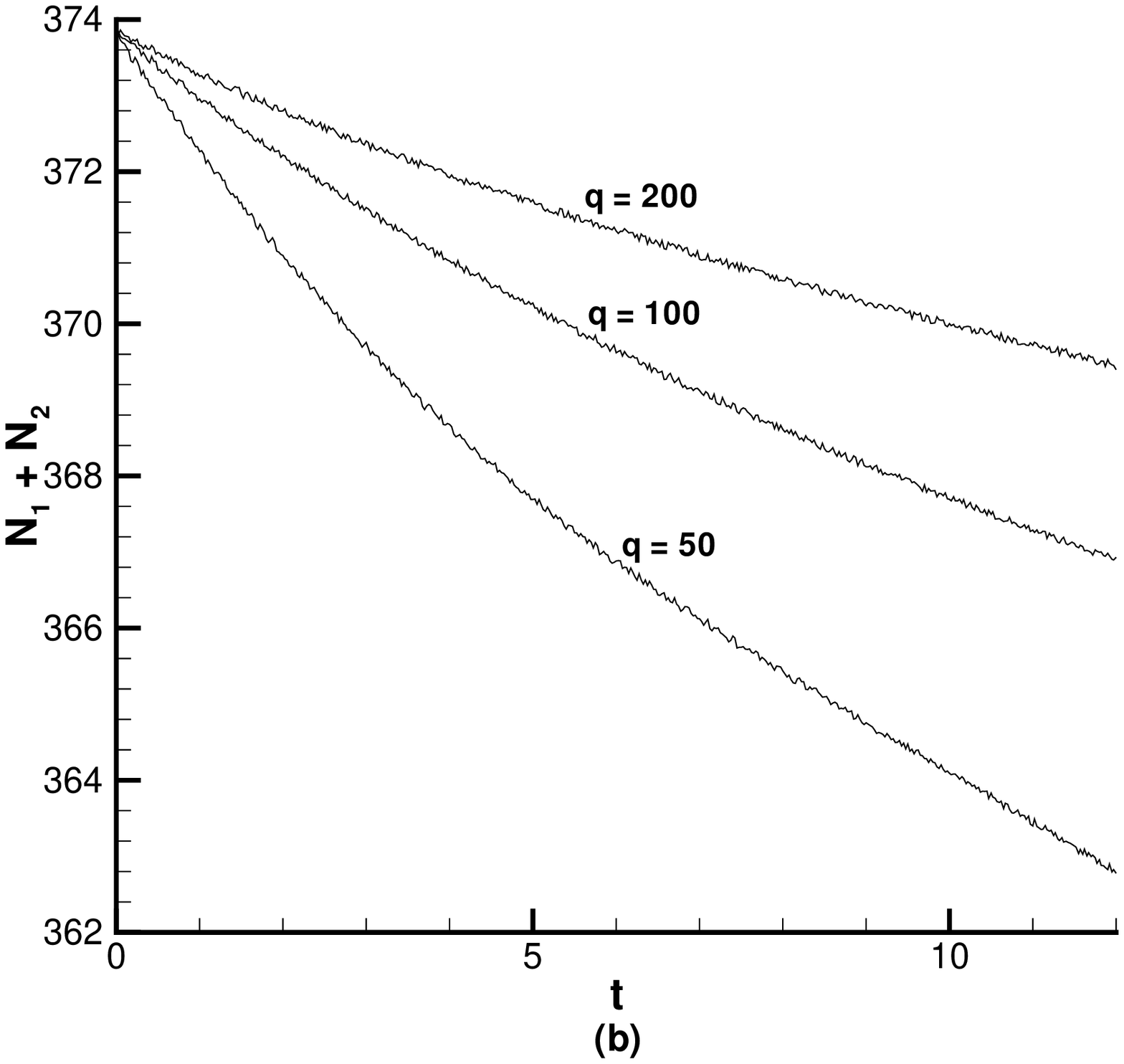}
\end{array}$\caption{The influence of the variation of the boundary-condition parameter
$q$ (the momentum of atoms in pulses released by the matter-wave laser), see
Eq. (\protect\ref{tau}), on the time dependence of the outcoupling rate (a),
and on the depletion dynamics (b). Except for $q$, other parameters are the
same as in Fig. \protect\ref{fig3}.}
\label{fig6}
\end{figure}

\begin{figure}[tbp]
$\begin{array}{cc}
\includegraphics[width=3.7in]{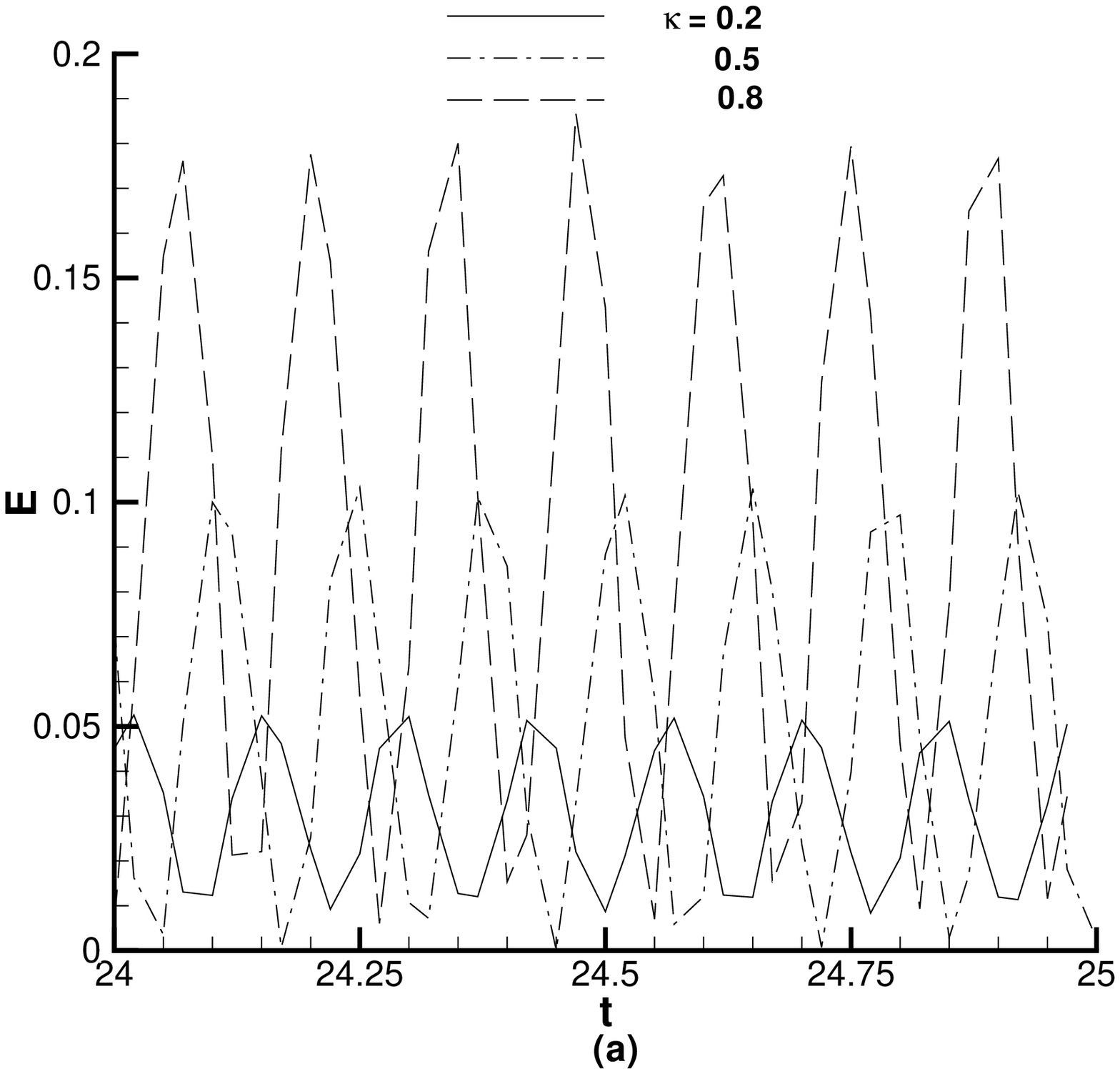} & \includegraphics[width=3.7in]{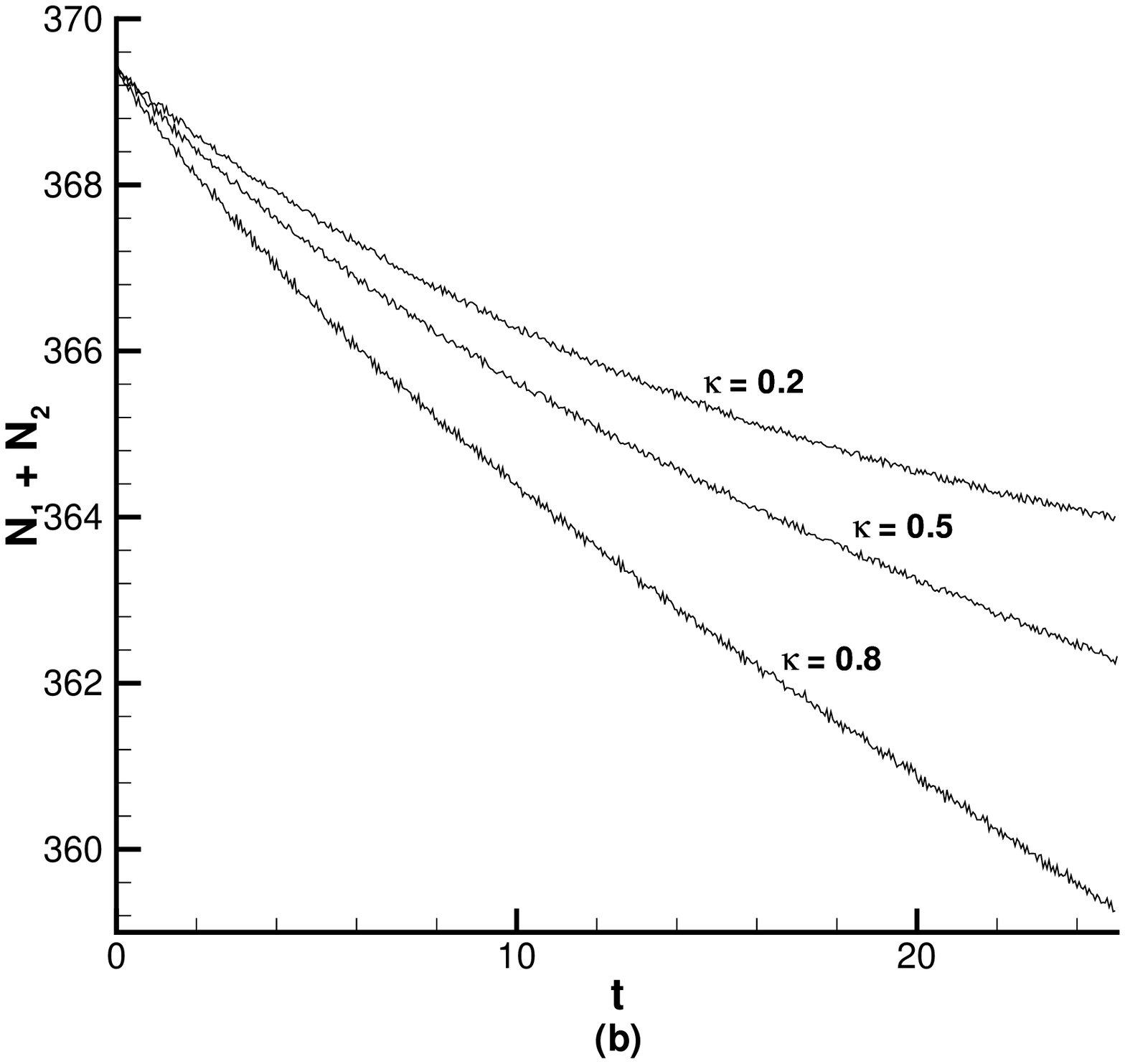}
\end{array}$\caption{The same as in Fig. \protect\ref{fig6}, but showing the effect of
the variation of the strength of the linear coupling between the reservoir
and cavity. }
\label{fig7}
\end{figure}

A more important role is played by the constant $q$ in Eq.
(\ref{tau}) (recall it determines the momentum of atoms in the
outcoupling pulse), and the linear-tunneling constant $\kappa $
that determines the strength of the linear coupling between the
reservoir and cavity in Eqs. (\ref{psi}) and (\ref{phi}). Figures
\ref{fig6} and \ref{fig7} illustrate effects of the variation of
these parameters on the operation regime. This is shown via the
change of the time dependences of the outcoupling rate $E(t)$ and
total amount of matter remaining in the system, $N_{1}+N_{2}$.

\begin{figure}[tbp]
$\begin{array}{cc}
\includegraphics[width=3.7in]{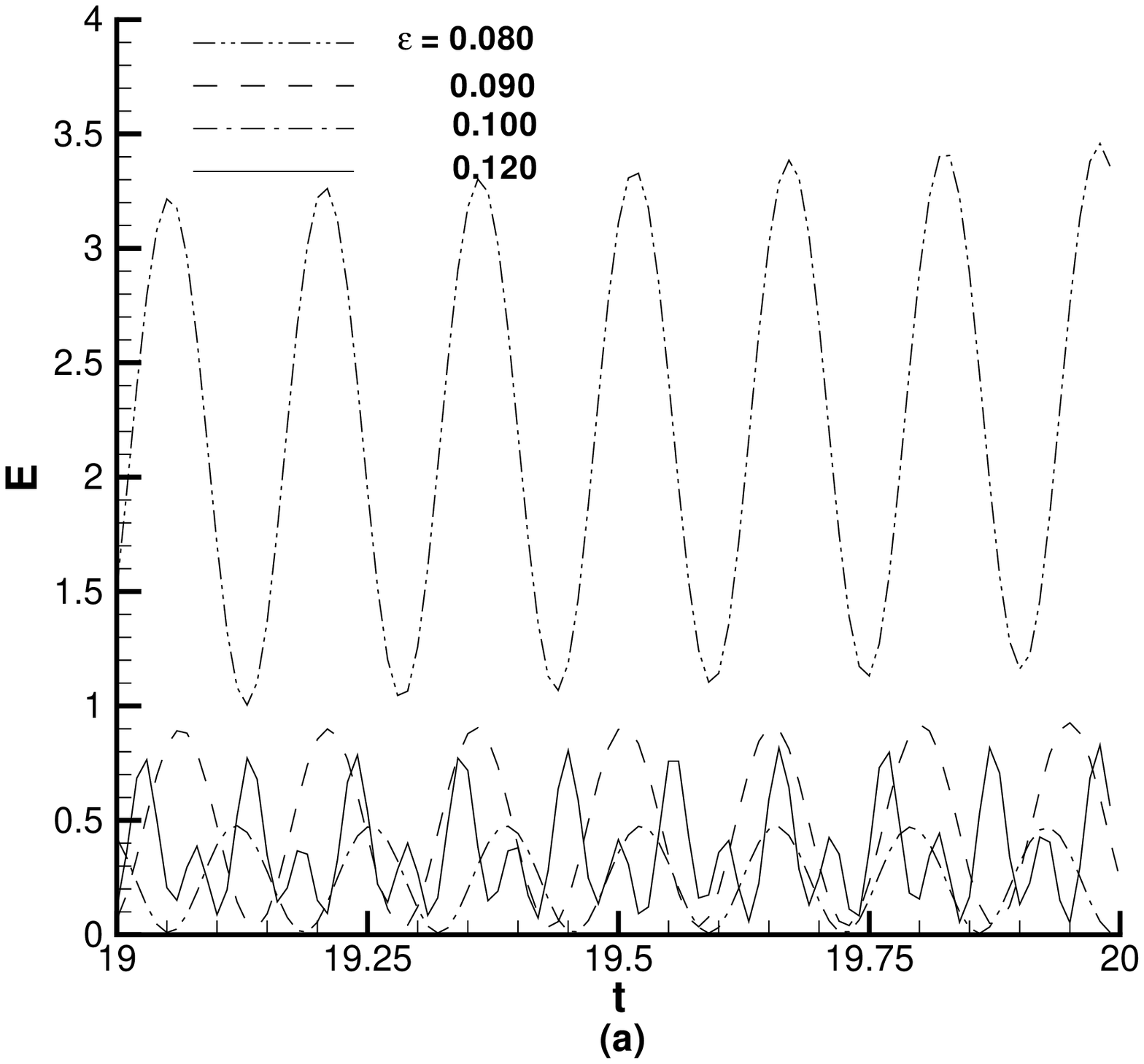} & \includegraphics[width=3.7in]{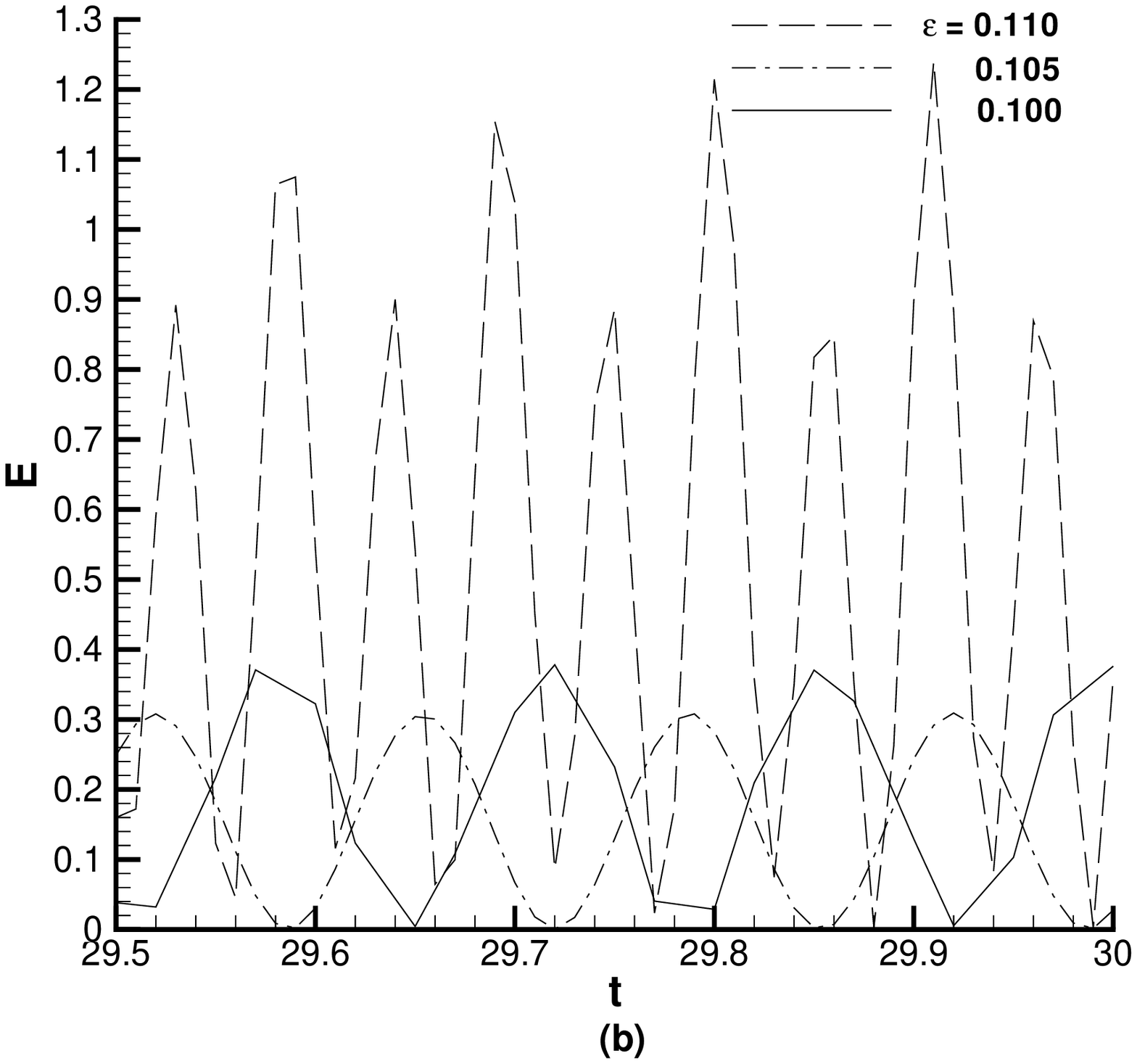}
\end{array}$\caption{The effect of variation of the nonlinearity coefficient
$\protect\epsilon $ in the reservoir on the operation regime of the matter-wave laser
model. Other parameters are the same as in Fig. 3.}
\label{fig8}
\end{figure}

The essential dependence of the dynamical regimes on the
parameters $q$ and $\kappa $ is quite natural. In addition to
this, and somewhat unexpectedly, the system demonstrates strong
sensitivity to the value of the nonlinearity coefficient $\epsilon
$ in Eq. (\ref{phi}), which is proportional to the positive atomic
collision length in the reservoir. In fact, stable operation
regimes, providing for the generation of very long periodic arrays
of the outcoupling pulses, are possible in a relatively narrow
interval of values of $\epsilon $, if the other parameters are
fixed. The dependence of the dynamical states on $\epsilon $ is
illustrated, by displaying segments of the established dependence
$E(t)$, in Fig. \ref{fig8}. Panel (a) in this figure shows that,
at $\epsilon =0.08$, the generated array of pulses has a high
amplitude, but with shallow dips between pulses, so that they are
not well separated. The same panel demonstrates that the optimum
shape of the array of well-separated pulses is achieved between
$\epsilon =0.09$ and $0.12 $ (which implies the positive
scattering length in the reservoir being $\simeq 10$ times smaller
than the absolute value of the negative scattering length in the
cavity). The situation in the latter interval is additionally
illustrated by a set of curves in panel (b) of Fig. \ref{fig8}.
Physically, the necessary fine-tuning of $\epsilon $ in the
reservoir may be achieved by means of the Feshbach resonance.

Undoing normalizations that cast the coupled GPEs in the rescaled form of
Eqs. (\ref{psi}) and (\ref{phi}), it is easy to derive a relation between
the actual number of atoms, $N_{\mathrm{phys}}$, and the norm $N_{1}$ of the
$\psi $ function in the normalized units:
\begin{equation}
N_{\mathrm{phys}}=\frac{r^{2}}{8|a|L}N_{1},  \label{Nphys}
\end{equation}where $r$ is the transverse radius of the cigar-shaped trap, which plays the
role of the cavity, $a$ and $L$ being, respectively (as defined above), the
Feshbach resonance-controlled atomic scattering length in the cavity and the
cavity's length, both taken in physical units. For the cases displayed in
Fig. \ref{fig8}, a characteristic value of the norm of an individual
released pulse, in the normalized units, is $N_{1}\sim 0.1$. Assuming the
aspect ratio of the cylindrical trap to be $L/(2r)\simeq 10$, and the actual
length $L\simeq 100$ $\mu $m, Eq. (\ref{Nphys}) yields the number of atoms
per outcoupling pulse between $10$ and $1000$, if the scattering length is
kept at a low level, between $a\simeq -0.1$ nm and $a\simeq -0.01$ nm (a
very small value of $|a|$ is necessary to prevent collapse in the cavity
trap). These values may be sufficient for physical applications. In
particular, the threshold number of $^{7}$Li atoms, necessary for the
creation of a stable three-dimensional solitons in an optical lattice, was
recently shown to be $60$ \cite{3D}. The number of atoms $\sim 100$ may also
be sufficient for direct optical detection of the matter-wave pulse. As for
the total number of atoms originally stockpiled in the reservoir, the above
estimates, including Eq. (\ref{cycles}), yield the range of $N_{0}\sim
10^{5}-10^{7}$ for it.

\section{Conclusion}

The objective of this work was to propose a model of a matter-wave
laser which can provide for stable periodic generation of a
sequence of solitary-wave pulses from a reservoir containing a
large amount of matter in the form of a coherent Bose-Einstein
condensate. The system includes two parallel cigar-shaped traps,
\textit{viz}., the reservoir and the work cavity, which are
coupled by the tunneling of atoms between them. The scattering
length of atomic collisions is tuned to be positive and negative
in the reservoir and cavity, respectively. Both ends of the
reservoir, and the left edge of the cavity are closed by lids.
Solitary pulses are released through a ``valve" at the right edge
of the cavity, which is described by the linear boundary condition
(\ref{tau}), and may be implemented as a potential step separating
the cavity and the outcoupling waveguide. Two different regimes of
the operation of the matter-wave laser were identified:
circulations of a narrow solitary pulse in the cavity, and
vibrations of a broad standing lump. Only the latter mode is
stable, while the circulation regime spontaneously rearranges
itself into the vibration mode. Dependence of the stability and
characteristics of the pulse-generating regime on parameters of
the model was explored. The regime is sensitive to the
boundary-condition parameter $q$ and strength $\kappa $ of the
linear coupling between the core and reservoir, and especially
sensitive to the value of the nonlinearity coefficient (i.e.,
scattering length) in the reservoir.

The bottom line is that the model can guarantee stable generation of a
pattern consisting of up to $10^{4}$ permanent-shape pulses, each containing
between $10$ and $1000$ atoms, under typical experimental conditions. It is
relevant to stress that the parameter space of the model, which is huge
(seven-dimensional, as the full set of the parameters includes $\epsilon
,g,f,\kappa ,q$, and the initial norms $N_{1,2}^{(0)}$), is far from being
fully explored, so still more promising regimes may be hidden in the model.

\section*{Acknowledgements}

We appreciate valuable discussions with J. Brand, L. Carr, and V.
M. P\'{e}rez-Garc\'{\i}a. The work of B.A.M. was partially
supported by the Israel Science Foundation through the
Excellence-Center grant No. 8006/03.

\end{document}